\documentclass{article}
\usepackage{GSMemorial}
\usepackage{pslatex}
\usepackage{graphicx}
\frompage{000} \topage{000}                                              
\newcommand{\nn}{\nonumber}
\newcommand{\be}{\begin{equation}}
\newcommand{\bel}[1]{\begin{equation} \label{#1} }
\newcommand{\ee}{\end{equation}}
\newcommand{\bea}{\begin{eqnarray}}
\newcommand{\beal}[1]{\begin{eqnarray} \label{#1} }
\newcommand{\eea}{\end{eqnarray}}

\def\rightmark{A Colorful Wake}
\def\leftmark{B. M\"uller and J. Ruppert}
                                                                               
\title{A Colorful Wake for Gerhard Soff}
\authors{
{Berndt M\"uller and J\"org Ruppert
\index{M\"uller, B.}
\index{Ruppert, J.}
}\\[2.812mm]
{\normalsize
\hspace*{-8pt}
Department of Physics, Duke University, Durham, NC 27708, USA
}}
  
\abstract{We calculate the wake induced in a hot QCD plasma by a fast 
parton in the framework of linear response theory. We discuss two 
scenarios: ($i$) a weakly coupled quark-gluon plasma described by 
hard-thermal loop perturbation theory and ($ii$) a strongly coupled 
quark-gluon plasma which resembles a quantum liquid. We show that a
Mach cone can appear in the second scenario, but not in the first one.}
                                                                               
\makeindex

\begin{document}

\maketitle 
\setcounter{page}{1}

\section{Introduction}\label{intro}

The Oxford English Dictionary records three different meanings of the 
noun {\em wake}:
\begin{itemize}
\item a trail of disturbed water or air left by the passage of a ship
      or aircraft;
\item a watch or vigil held beside the body of someone who has died;
\item (especially in Ireland) a party held after a funeral.
\end{itemize}
Since the Greek word $\sigma\upsilon\mu\pi\omicron\sigma\iota\omicron\nu$ 
describes a gathering of friends to eat, drink, and converse, 
the {\em Symposium} in memory of our deceased friend and 
colleague Gerhard Soff certainly fits the third definition. 
My lecture will be mainly concerned with the first definition, in a 
slightly generalized sense. To define the appropriate context, we first 
need to review some of the salient results from the experiments 
conducted at the Relativistic Heavy Ion Collider (RHIC).

The quenching of QCD jets in relativistic heavy ion collisions due to the
energy loss suffered by hard partons as they traverse dense matter was 
proposed more than two decades ago as an important indicator for the 
creation of a quark-gluon plasma 
\cite{Bjorken:1982tu,Gyulassy:1990ye,Wang:1991xy}. Over the past five 
years, this phenomenenon has been extensively studied experimentally 
at RHIC. One looks for hadrons with a high transverse momentum $p_T$,
which are produced when an energetic, hard scattered parton fragments
into hadrons far outside the remnants of the nuclear collision. Since
both, the scattering probability and the fragmentation probability, are 
known or calculable in QCD, the sole unknown is the amount of energy 
lost by the parton before it fragments into hadrons. This makes it 
possible to deduce the energy loss from the measured hadron yield.

The RHIC data clearly demonstrate a strong suppression of the emission 
of high-$p_T$ hadrons in Au + Au collisions. The suppression is found to grow 
in severity with increasing centrality. The suppression factor $R_{AA}$,
defined as the ratio of the hadron yield in Au + Au collisions compared 
to the yield in $p+p$ collisions scaled by the 
appropriate number of binary nucleon-nucleon interactions in the nuclear
collision, reaches about 1/5 in the most central events \cite{Adcox:2001jp}. 
Most of the observed hadrons originate near the surface of the interaction
region oriented toward the detector. Hadrons from the companion jet emitted
in the opposite direction, which has a much longer path through the medium,
are even more strongly suppressed \cite{Adler:2002tq}.

The main emphasis in theoretical studies of jet quenching has been on the 
description of the energy loss which the leading parton suffers due to the 
emission of a secondary partonic shower when traversing the medium. Reviews
of the theory of medium induced energy loss and its associated phenomenology
are found in \cite{Baier:2000mf,Kovner:2003zj,Accardi:2003gp,Jacobs:2004qv}. 
The main result is that the energy loss of a hard parton in dense QCD matter
is dominated by radiative processes involving gluon emission after elastic
collisions of the parton with color charges (mostly gluons) contained in the 
medium. Coherence effects lead to a quadratic dependence of the total energy 
loss on the traversed path length $L$. The stopping power of the medium 
is encoded in the quantity
\bel{eq01}
\hat q = \rho \int q^2 \frac{d\sigma}{dq^2} dq^2 ,
\ee
where $\rho$ denotes the gluon density of the medium and $d\sigma/dq^2$ is
the differential cross section for elastic scattering.

An interesting question is: What happens to the radiated energy? Or more
generally: How does the medium respond to the penetration by the hard parton?
The RHIC experiments are just beginning to address this intriguing question. 
First indications of a medium response have been seen in three sets of data:
\begin{enumerate}
\item
Although energetic hadrons in the direction opposite to a high-$p_T$ trigger 
hadron are almost completely suppressed in central collisions 
\cite{Adler:2002tq}, one finds an increased yield of soft, low-$p_T$ hadrons 
\cite{Adams:2004gp,Wang:2004kf}.
\item
The angular distribution of the soft hadrons emitted in the direction opposite
to an energetic hadron does not seem to peak at $180^{\circ}$ but at a 
smaller angle \cite{Rak:2005st,Adler:2005ee}.
\item
Correlated two-hadron emission in the same direction in the momentum range 
0.15 GeV/c $< p_T <$ 2 GeV/c is enhanced \cite{Adams:2004pa}.
\end{enumerate}

Since the quark-gluon plasma is a plasma, after all, it is natural to begin
a theoretical investigation of the medium response by applying the tools 
which have been successfully used to describe the response of a metallic
electron plasma to the penetration by a fast ion. Such a charged projectile
induces a wake of charge and current density in the target, accompanied by
induced electric and magnetic fields. The wake, which has the shape of a
Mach cone, reflects significant aspects of the response of the medium.

After a brief review of the evidence for this phenomenon in condensed matter
physics, we apply the same methods of linear response theory to the system 
of a relativistic color charge traveling through a QCD plasma. We calculate 
the plasma reponse to an external point source traveling at a velocity close 
to the speed of light. In this framework, quantum effects are included 
implicitly via the dielectric functions, $\epsilon_L$ and $\epsilon_T$. 
For simplicity, we shall assume that the medium is homogeneous and isotropic, 
and we disregard finite size effects. We then consider two models for the
color response of the QCD plasma: ($i$) the response predicted by QCD
perturbation theory, and ($ii$) a response of the kind expected for a 
strongly coupled, ``liquid'' plasma. We present the results of our 
calculations of the wake structure behind a fast color charge for both
scenarios and discuss the general conditions which must be met for the
wake to have a Mach cone-like structure \cite{Ruppert:2005uz,Ruppert:2005sj}.

\section{Plasma linear response theory}

The linear response of the plasma to an external electromagnetic field has 
been extensively studied in plasma physics (see e.~g.~\cite{Ichimaru}). In 
this formalism, a dielectric medium is characterized by the components of 
the  dielectric tensor $\epsilon_{ij}(\omega,k)$. For an isotropic, 
homogenous medium the dielectric tensor can be decomposed into its 
longitudinal and transverse components characterized by the dielectric 
functions $\epsilon_L(\omega,k)$ and $\epsilon_T(\omega,k)$:
\bel{eq02}
\epsilon_{ij}=\epsilon_L {\cal P}_{L,ij} +\epsilon_T {\cal P}_{T,ij}.
\ee
Here ${\cal P}_{L,ij}=k_i k_j/ k^2$ and ${\cal P}_T=1-{\cal P}_L$ are the 
longitudinal and transverse orthonormal projectors with respect to the 
momentum vector $\vec{k}$. One can relate the dielectric functions 
$\epsilon_L$ and $\epsilon_T$ to the self-energies $\Pi_L$ and $\Pi_T$ 
of the in-medium photon via \cite{LeBellac}:
\bel{eq03}
\epsilon_L(\omega,k)=1-\frac{\Pi_L(\omega,k)}{\omega^2-k^2}, 
\qquad\qquad
\epsilon_T(\omega,k)=1-\frac{\Pi_T(\omega,k)}{\omega^2} .
\ee
Using Maxwell's equations and the continuity equation in momentum space, 
the total electric field $\vec{E}_{\rm tot}$ in the plasma is related to 
the external current $\vec{j}_{\rm ext}$ via:
\beal{EtotTOJ}
\left[\epsilon_L {\cal P}_{L} + \left(\epsilon_T -\frac{k^2}{\omega^2}\right) 
  {\cal P}_{T}\right] \vec{E}_{\rm tot}(\omega,k)
=\frac{4 \pi}{i \omega} \vec{j}_{\rm ext} (\omega,k) .
\eea
Equation (\ref{EtotTOJ}) has propagating solutions when the determinant 
constructed from the elements of the tensor vanishes:
\bel{eq05}
{\rm det}\left|\epsilon_L {\cal P}_{L} + \left(\epsilon_T -\frac{k^2}{\omega^2}\right) 
  {\cal P}_{T}\right|=0 .
\ee
This equation governs the dispersion relation for the waves in the medium. 
It can be diagonalized into purely longitudinal and transverse parts, yielding
dispersion relations for the longitudinal and transverse dielectric functions 
\cite{Ichimaru}:
\bel{Dispersion}
\epsilon_L(\omega,k) = 0, 
\qquad\qquad
\epsilon_T(\omega,k) = (k/\omega)^2.
\ee
These equations determine the longitudinal and transverse plasma modes. 
The longitudinal equation is the dispersion relation for density 
fluctuations in the plasma, namely space-charge fields which can 
propagate through the plasma without far away from an external 
perturbation. 

The charge density induced in the wake by the external charge distribution is:
\bel{charge}
\rho_{\rm ind}=\left(\frac{1}{\epsilon_L}-1\right)\rho_{\rm ext}.
\ee
In the transverse gauge, the induced charge density is related to the 
induced Coulomb potential via the Poisson equation:
$k^2\phi_{\rm ind}= 4\pi\rho_{\rm ind}$. Since one can relate the total 
electric field to the induced charge in linear response theory by 
\bel{eq09}
\vec{j}_{\rm ind}=\frac{i\omega}{4\pi} (1-\epsilon)\vec{E}_{\rm tot} ,
\ee
a direct relation between the external and the induced current can be \
derived using (\ref{EtotTOJ}):
\bel{current}
\vec{j}_{\rm ind}=\left[\left(\frac{1}{\epsilon_L}-1\right){\cal P}_L 
  + \frac{\omega^2(1-\epsilon_T)}{\omega^2\epsilon_T-k^2} 
    {\cal P}_T\right]\vec{j}_{\rm ext}.
\ee
The induced charge and the induced current obey the continuity equation:
\be \label{continuity}
i\vec{k}\cdot\vec{j}_{\rm ind}-i\omega \rho_{\rm ind}=0.
\ee

Finally, we need to specify the form of the external current. For a 
fully stripped ion (or a single energetic parton) it is appropriate
to assume the current and charge density of a point-like charge moving 
along a straight line trajectory with constant velocity $\vec v$, 
whose Fourier transform is given by \cite{Neufeld}:
\beal{eq11}
\vec{j}_{\rm ext} &=& 2\pi q \vec{v} \delta(\omega-\vec{v} \cdot \vec{k}),
\nn \\
\vec{\rho}_{\rm ext} &=& 2\pi q \delta(\omega-\vec{v} \cdot \vec{k}) .
\eea

All equations given above, from (\ref{eq02}) to (\ref{eq11}), immediately
generalize to QCD by the simple addition of a color index $a=1,\ldots,8$ 
to the charge, the current, and the field strength, which are all in the 
adjoint representation of color SU(3) :
\be
q\to q^a, \qquad
(\rho,\vec j) \to (\rho^a,\vec j^a), \qquad
\vec E \to \vec E^a .
\ee
Since we are limiting the treatment of the medium response to effects 
that are linear in the perturbation, all nonlinear terms arising from 
the nonabelian nature of the color field are discarded.
The strength of the color charge of the projectile is defined by 
$q^a q^a = C_2 \alpha_s$ with the strong coupling constant 
$\alpha_s=g^2/4\pi$ and the quadratic Casimir invariant $C_2$ 
($C_F=4/3$ for a quark or antiquark and $C_A=3$ for a gluon). In this 
linearized treatment one disregards changes of the color charge while the 
particle is propagating through the medium. The nonabelian character of 
the QCD plasma only enters indirectly via the chromo-dielectric functions 
$\epsilon_{L/T}(\omega,k)$, which are scalars in color space
\cite{Weldon:1982aq,Klimov:1982bv,Thoma:1990fm}. Due to the 
self-interacting nature of the gluon in SU(3), gluons contribute to the 
polarization of the medium; in fact, they make the largest contribution.
 
The non-radiative part of the energy loss of the incident (color) charge 
is given by the back reaction of the induced (chromo-)electric field onto 
the incident particle. The energy loss per unit length is given by 
\cite{Ichimaru}:
\beal{energy1}
\frac{dE}{dx}
=q \frac{\vec{v}}{v}\,{\rm Re}\,\vec{E}_{\rm ind}(\vec{x}=\vec{v}t,t),
\eea
where the induced electric field $\vec{E}_{\rm ind}$ is the total electric 
field minus the vacuum contribution. Using the inverse of (\ref{EtotTOJ}), 
the induced field is given by:
\bel{loss}
\vec{E}_{\rm ind}
= \left[\left(\frac{1}{\epsilon_L}-1\right){\cal P}_L 
    + \left(\frac{\omega^2}{\omega^2\epsilon_T-k^2}
    - \frac{\omega^2}{\omega^2-k^2} \right){\cal P}_T\right] 
  \frac{4\pi}{i\omega}\vec{j}_{\rm ext} .
\ee
From (\ref{energy1}) and ({\ref{loss}) the non-radiative energy loss per 
unit length is given by \cite{Thoma:1990fm}:
\bel{energy2}
\frac{dE}{dx} = -\frac{C \alpha_s}{2 \pi^2 v} \int d^3k 
  \frac{\omega_k}{k^2}\,\left[{\rm Im}\,\epsilon_L^{-1}  
  + (v^2 k^2 - \omega_k^2)\,{\rm Im}\,(\omega_k^2\epsilon_T^{-1}- k^2)^{-1}
    \right] ,
\ee 
where $\omega_k=\vec{v}\cdot\vec{k}$.

\section{Quick flashback: Electron wakes in metal foils}

The formation of electromagnetic wakes induced by a fast ion in the 
electron plasma of a thin metal foil was investigated starting around 1980 
by Groeneveld and collaborators at Frankfurt \cite{Groeneveld1,Groeneveld2}. 
Their study was motivated, in part, by calculations 
\cite{Schafer:1978jq,Schafer:1980bg} done by a graduate
student, Wolfgang Sch\"afer,\footnote{Before doing this work for his
doctoral thesis, Wolfgang had worked under Gerhard Soff's supervision
calculating the influence of the vacuum polarization potential on the 
motion of two colliding heavy nuclei.} who had solved the equations 
of the previous section for a dielectric function of the Bloch type
\cite{Bloch} 
\bel{Bloch}
\epsilon_L = 1+\frac{\omega_p^2}{u^2 k^2-\omega^2} \quad  (k\le k_c) ,
\ee
where $\omega_p$ is the plasma frequency and $u$ denotes the sound
velocity in the plasma. The calculations showed that the wake in the 
electron plasma has the form of a Mach cone, similar to the Mach shock 
phenomena predicted for relativistic nucleus-nucleus collisions
\cite{Glassgold,Scheid:1973}. In our publications, we not only showed the 
detailed shape of the spatial distribution of induced plasma charge 
and current (see Fig.~\ref{figureMach}), but we also predicted the 
angular distribution of the electrons which would be emitted when the 
current wave hits the surface of the metal foil (see Fig~\ref{figureMach}). 
The peak of this distribution roughly coincides with the ``Mach angle'', 
given by
\bel{Mach}
\varphi_{\rm M}={\rm arccos}(u/v) .
\ee

\begin{figure}
\begin{center}
\includegraphics[width=0.45\textwidth]{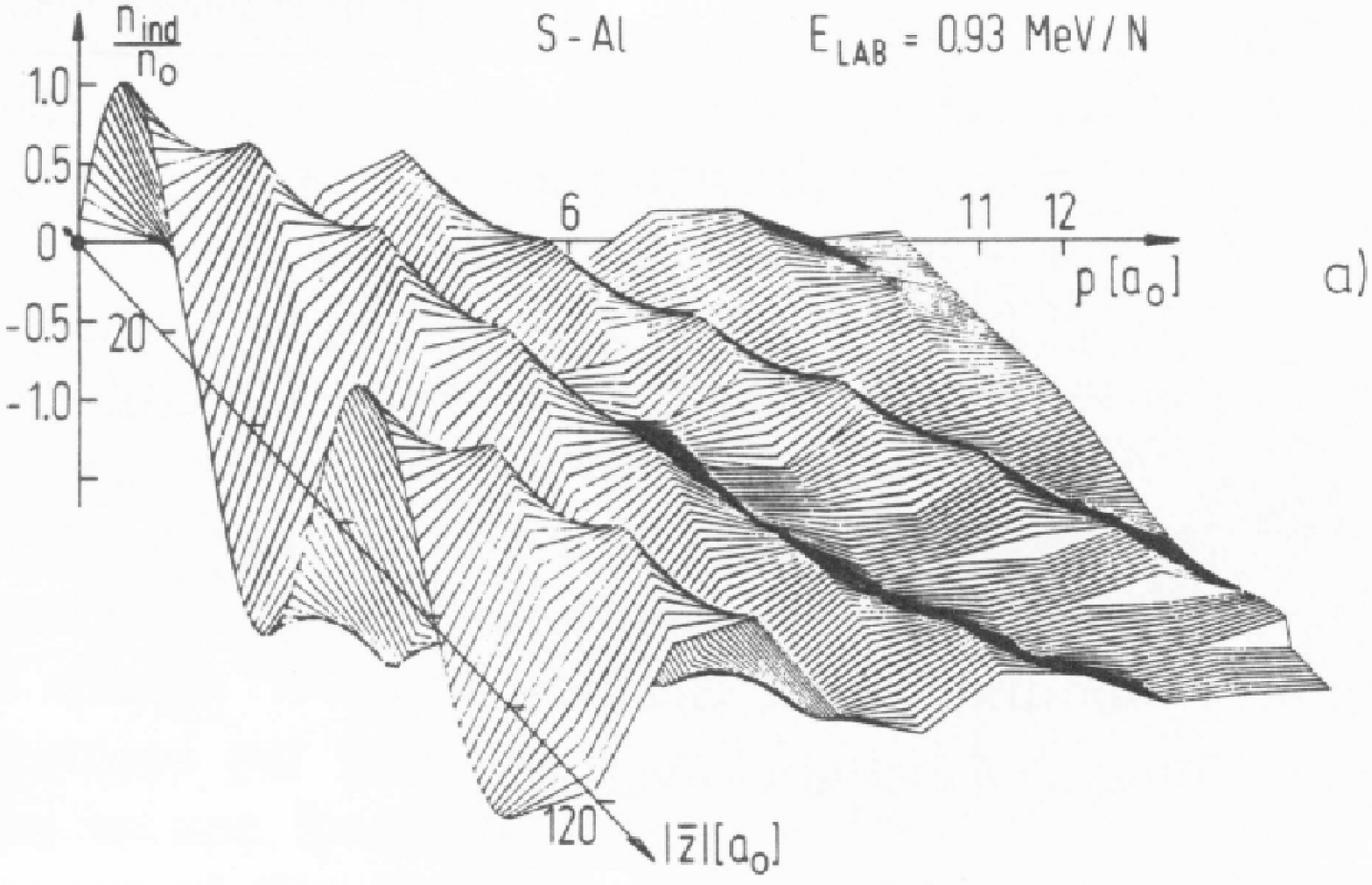}
\hspace{0.05\linewidth}
\includegraphics[width=0.45\textwidth]{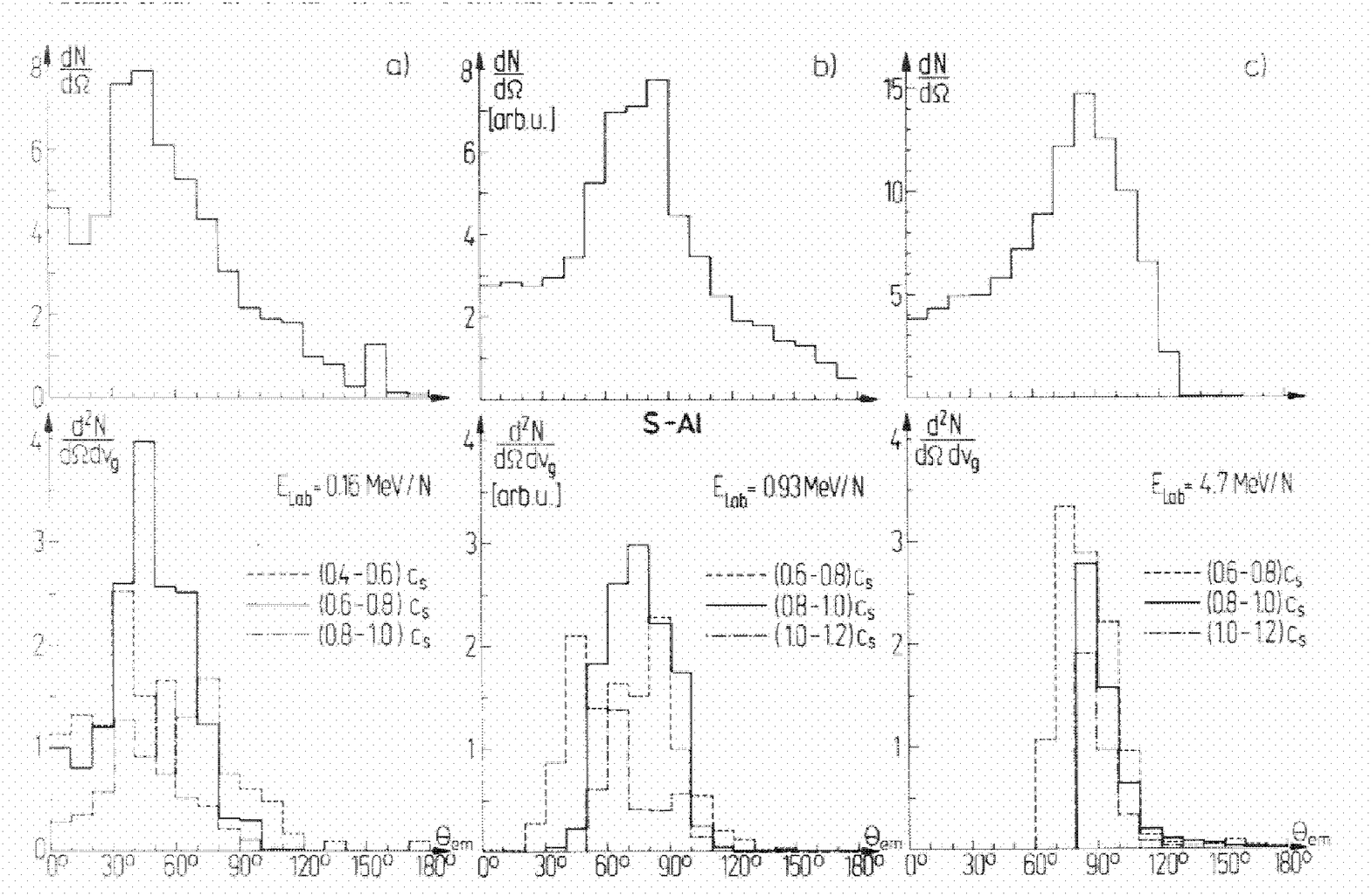}
\caption{Left: Spatial distribution of the induced charge density behind 
         an ion traveling through a metal. The plot represents a radial
         cut through the Mach cone. 
         Right: Predicted angular distribution of the electrons contained 
         in the wake current.
\label{figureMach}}
\end{center}
\end{figure}

The first published data confirmed the presence of directed
electron emission, and the dependence of the peak angle on the beam 
energy qualitatively agreed with our predictions \cite{Groeneveld1}.
However, it turned out that the results were not well reproducible
unless the surface of the foil was extremely clean. Groeneveld's
team therefore later repeated the measurement with foils whose surface
had been sputter-cleaned. The new data (see Fig.~\ref{figGroene}) showed 
much more pronounced and reproducible peaks \cite{Groeneveld2}. Their 
positions as a function of beam energy and Fermi energy of the target 
nicely followed the Mach relation (\ref{Mach}) in agreement with the 
predictions \cite{Schafer:1980bg}.

The experimentalists continued to study many aspects of this phenomenon 
in great detail, including the energy spectra and angular distributions 
of the ejected electrons, the refraction of the Mach wave at the planar 
surface of the foil, and the response of high $T_c$ superconductors 
\cite{Groeneveld3}. In recent years, the collective plasma waves excited 
by fast ions in soft biological tissue have been considered as a mechanism 
that contributes to the damage to living cells by fast C$^{6+}$ ions
\cite{Rothard}, which is an important factor in cancer therapy with 
heavy ion beams \cite{Cancer}.

\begin{figure}
\begin{center}
\includegraphics[width=0.4\textwidth]{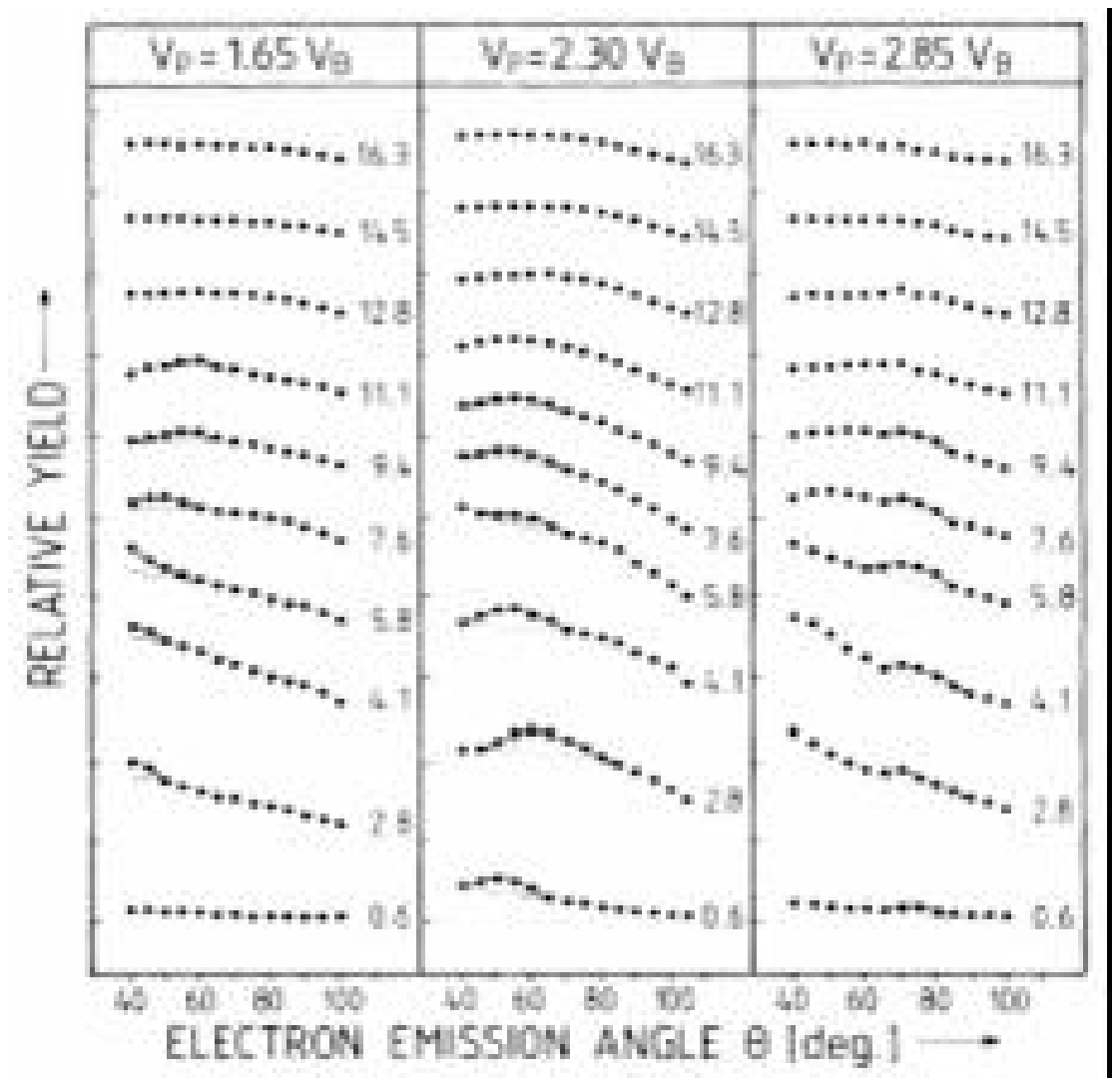}
\hspace{0.05\linewidth}
\includegraphics[width=0.5\textwidth]{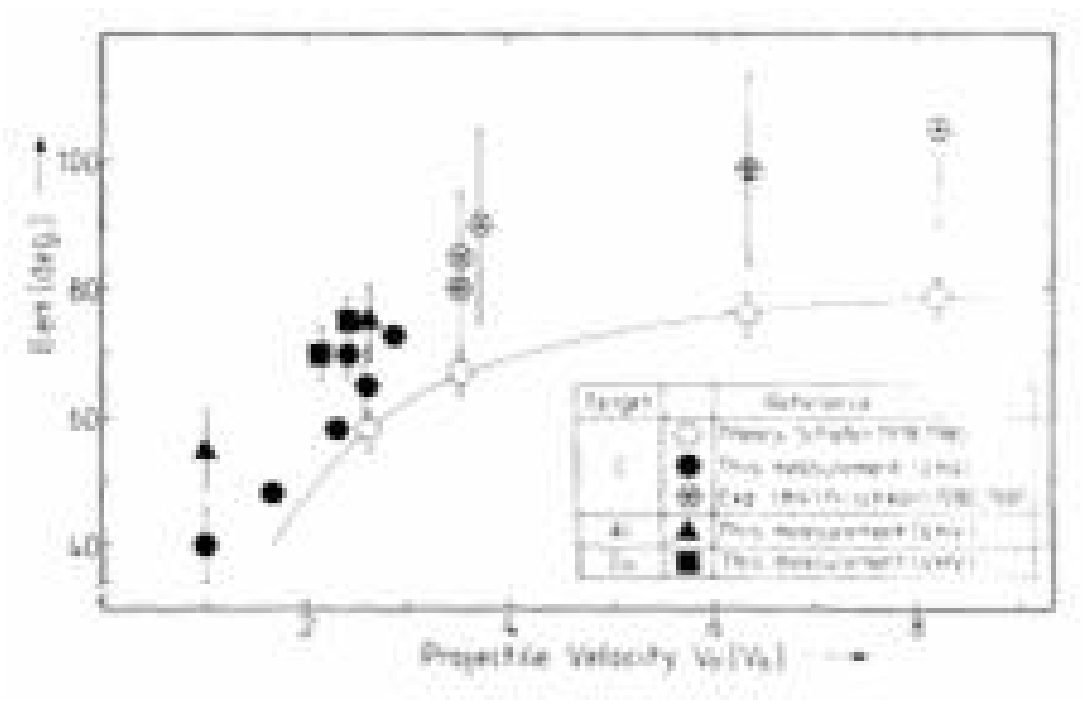}
\caption{Left: Angular distribution of electrons emitted after bombardment
         of thin metal foils by fully stripped ions. Right: Dependence of
         the peak position in the angular distribution for various target
         materials (C, Al, Cu) and different beam energies, in comparison
         with the predicted Mach angle for a carbon target 
         (from \protect\cite{Groeneveld2}).
\label{figGroene}}
\end{center}
\end{figure}

\section{Color wake in the high temperature approximation}
 
We now return to the problem of interest to us: the medium response to an
energetic parton. We shall discuss two qualitatively different scenarios. 
We first assume that the plasma is in the weakly coupled high temperature 
regime, where the gluon self-energy can be described by the leading order 
of the high temperature expansion, $T\gg \omega,k$, commonly called the 
hard-thermal loop (HTL) approximation \cite{Pisarski:1989cs,Braaten:1989mz}. 
This regime can be expected to be realized far above the deconfinement 
temperature $T_c$. 
The dielectric functions read \cite{Klimov:1982bv,Weldon:1982aq}:
\beal{eps}
 \epsilon_L &=& 
  1+\frac{2m_g^2}{k^2} \left[ 1-\frac{1}{2}x \left(
  {\rm ln}\left|\frac{x+1}{x-1}\right|-i\pi \Theta \left(1-x^2\right)
  \right) \right], 
\\
 \epsilon_T &=& 
  1-\frac{m_g^2}{\omega^2} \left[ x^2+ \frac{x(1-x^2)}{2} 
 \left( {\rm ln}\left|\frac{x+1}{x-1}\right|
  -i\pi \Theta \left( 1-x^2 \right) \right) \right],
\eea
where $x=\omega/k$. We first discuss the induced charge and current densities 
$\rho_{\rm ind}$ and $\vec{j}_{\rm ind}$. The Fourier transform of 
(\ref{charge}) in cylindrical coordinates is given by:
\beal{charge2}
\rho^a_{\rm ind}(\rho, z, t)
&=& \frac{m_g^3 q^a}{(2 \pi)^2 v} \int_0^\infty d\kappa' \kappa' 
    J_0(\kappa\rho) \times 
\\ &&
    \int^{\infty}_{-\infty} d\omega' \,
    {\rm exp} \left[i\omega\left(\frac{z}{v}-t\right)\right] 
    \left(\frac{1}{\epsilon_{L}}-1 \right),
\nn
\eea
where $k=\sqrt{\kappa^2+\omega^2/v^2}$ and $\omega=m_g\omega'$,
$\kappa=m_g\kappa'$. 
This shows that the induced charge density $\rho^a_{\rm ind}$ is 
proportional to $m_g^3$. The cylindrical symmetry around the jet axis 
restricts the form of the current density vector $\vec{j}^a_{\rm ind}$. 
It only has non-vanishing components parallel to the beam axis: 
\beal{current2a}
j^a_{\rm v, ind}(\rho, z,t) &=&
  \frac{m_g^3 q^a}{(2 \pi)^2 v^2} \int_0^\infty d\kappa' \kappa' 
  J_0(\kappa\rho) 
\\ &&
  \int^{\infty}_{-\infty} d\omega' {\rm exp} 
  \left[i\omega\left(\frac{z}{v}-t\right)\right] 
  \left[\left(\frac{1}{\epsilon_{L}}-1\right) \frac{\omega^2}{k^2} 
  + \frac{1-\epsilon_{T}}{\epsilon_{T}-\frac{k^2}{\omega^2}}
  \left(v^2-\frac{\omega^2}{k^2}\right) \right] , 
\nn 
\eea
and radially perpendicular to the beam axis:
\beal{current2b}
j^a_{\rho,\rm ind}(\rho, z ,t) &=& 
  \frac{i m_g^3 q^a}{(2 \pi)^2 v} \int_0^\infty d\kappa' \kappa' 
  J_1(\kappa\rho) 
\\ &&
  \int^{\infty}_{-\infty} d\omega' {\rm exp} 
  \left[i\omega\left(\frac{z}{v}-t\right)\right] 
  \frac{\omega \kappa}{k^2}\left[\left(\frac{1}{\epsilon_L}-1\right)
  -\left(\frac{1-\epsilon_T}{\epsilon_T-\frac{k^2}{\omega^2}}\right)\right] .
\nn
\eea
Again, the components of the current density are proportional to $m_g^3$. 

For the dispersion relations (\ref{eps}), longitudinal and transverse plasma 
modes can only appear in the time-like sector of the $\omega,k$ plane 
\cite{LeBellac,Weldon:1982aq}. Therefore, collective excitations do not 
contribute to the charge and current density profile of the wake. Emission 
analogous to Cherenkov radiation and Mach cones are absent, but the charge 
carries a screening color cloud along with it. Fig.~\ref{figure1}, which 
shows the charge density of a colored parton traveling with $v=0.99 c$, 
illustrates this physically intuitive result.

\begin{figure}
\centerline{\includegraphics[width=1\linewidth]{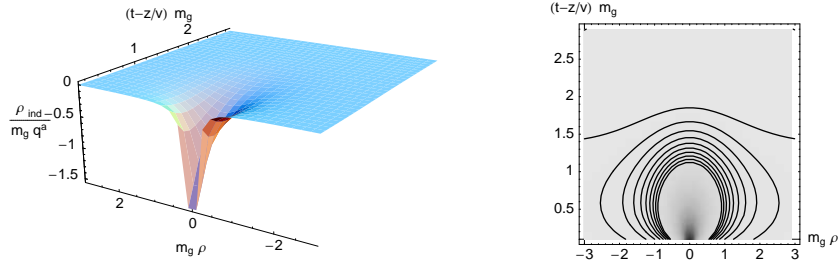}}
\caption{Spatial distribution of the induced charge density for a color
         charge traveling with velocity $v/c=0.99$ in a high temperature 
         QCD plasma where the HTL approximation applies. The left plot 
         shows equi-charge density lines in the rest frame of the charge.
\label{figure1}}
\end{figure}
 
In spite of the absence of a Mach cone, the particle loses energy due to 
elastic collisions in the medium, which can be described by formula 
(\ref{energy2}). This mechanism of energy dissipation has been studied 
in \cite{Thoma:1990fm}. The integrand in (\ref{energy2}) contributes to the 
integral in the space-like region only, where $|x|<1$, and hence does not 
get contributions from frequencies where collective plasma modes exist. 
This is consistent with the fact that such modes are not excited by the
moving color charge.

\section{Charge wake induced in a strongly coupled QGP}

In the second scenario we investigate what happens if the QCD plasma 
is a strongly coupled quark-gluon plasma (sQGP) that can be described as 
a quantum liquid. Our consideration of the second scenario is motivated 
by the RHIC experimental results on collective flow, which have led 
to the conclusion that the QCD plasma behaves like a nearly ideal 
fluid with very low viscosity (see e.~g.\ \cite{Teaney:2003pb}). 
This implies that long wavelength collective modes are almost undamped, 
while the short-distance dynamics
are strongly dissipative due to large transport cross sections. In
this respect, the sQGP resembles the electron plasma in metallic targets 
described in the previous section. The idea that a fast parton could 
excite Mach waves in such a QCD plasma and that the sound velocity of 
the expanding plasma could be determined from the emission pattern of 
the secondary particles traveling at an angle with respect to the jet 
axis was first suggested by St\"ocker \cite{Stoecker:2004qu}. The recent 
work by Casalderrey-Solana and Shuryak \cite{Casalderrey-Solana:2004qm}
explores the formation of a conical flow by a hard parton in a quark-gluon 
plasma within a hydrodynamical framework, but does not specify how the 
energy of the quenched jet is deposited into the medium. 

There are presently no theoretical methods available for first principle 
calculations of the color response functions in a strong coupled QGP plasma. 
We therefore confine our investigation to a simple model, which encodes
the essential differences between a quantum liquid scenario and the weak 
coupling scenario of the last section. The most prominent difference is 
the possibility that a plasmon mode may extend into the space-like region 
of the $\omega-k$ plane above some threshold value $k_s$. As we already 
emphasized, the sQGP paradigm suggests very low dissipation at small $k$, 
but large dissipation at high $k$. Our assumption is that a critical momentum 
$k_c$ separates the regimes of collective and single particle excitation modes 
in the quantum liquid where the dominant colored modes below $k_c$ are plasmon 
excitations with negligible dissipative single-particle coupling. Since we 
are here predominantly interested in collective effects in the plasma, we 
restrict our study to the region $k<k_c$ and simply cut off all Fourier 
integrals at $k_c$.

\begin{figure}
\centerline{\includegraphics[width=0.4\linewidth]{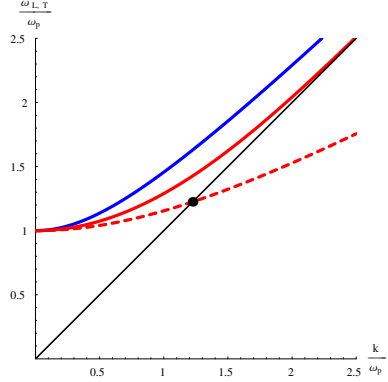}}
\caption{Dispersion relation for the plasmon mode for a weakly coupled 
         plasma described by the HTL formalism (solid red line) and for 
         a strongly coupled plasma described by the modified Bloch formula 
         (\ref{NonBloch}). The latter extends into the space-like region
         ($k>\omega$) of the $\omega-k$ plane. The dispersion curve of
         transverse HTL-plasma modes is shown in blue and the light cone
         ($\omega=k$) is represented in black. The black dot indicates the
         point $k_s$ where the dispersion curve intersects the light cone.
\label{figureDisp}}
\end{figure}

To be specific, we assume that the dielectric function of the strongly 
coupled plasma in the $k<k_c$ regime leads to a longitudinal dispersion 
relation of the form: 
\bea \label{Dispersion2}
\omega_{\rm L}=\sqrt{u^2 k^2 + \omega_p^2}\,\,,
\eea
where $\omega_p$ denotes the plasma frequency and $u<c$ is the speed of 
plasmon propagation, here assumed to be constant. In accordance with 
(\ref{Dispersion2}) we posit the following dielectric function
\bel{NonBloch}
\epsilon_L = 1+\frac{\omega_p^2/2}{u^2 k^2 - \omega^2 + \omega_p^2/2}
\qquad (k\le k_c) ,
\ee 
which differs from the classical, hydrodynamical dielectric function 
of Bloch \cite{Bloch} by remaining regular in the limit $k,\omega \to 0$. 
The Bloch function is singular at small $k,\omega$ due to the mixing of
the plasmon mode with the phonon mode. Such a mixing cannot occur in the
QCD plasma, because the plasmon and phonon belong to different irreducible 
representations (octet versus singlet) of color SU(3) and because the 
medium is charge symmetric. 

\begin{figure}
\centerline{\includegraphics[width=1\linewidth]{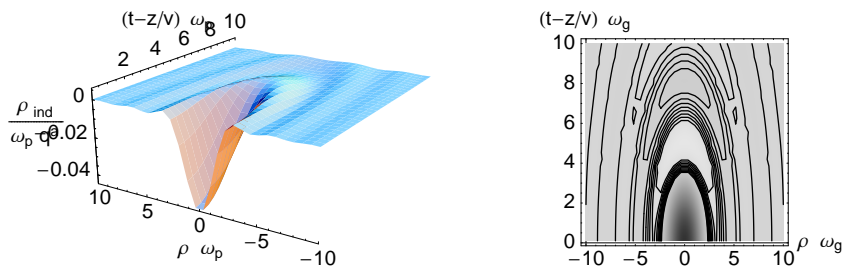}}
\caption{Spatial distribution of a induced charge density around a color
         charge traveling with velocity $v=0.55c<u$. The left plot shows 
         equi-charge density lines. The density profile is similar to that 
         of the cloud surrounding a color charge at rest.
\label{figure2}}
\end{figure}

The dielectric function (\ref{NonBloch}) is constructed in such a way that 
it allows us to study one specific aspect of a quantum liquid scenario: 
that the plasmon mode may extend into the space-like region of the $\omega-k$ 
plane. This behavior is illustrated in Fig.~\ref{figureDisp} by the dashed
red line, which cuts through the light cone at $k_s$ (at the black dot in
Fig.~\ref{figureDisp}) and continues into the space-like region. 
This contrasts with the dispersion relation of the longitudinal plasmon 
mode in the HTL formalism (solid red line), which always remains in the 
time-like region $\omega(k)>k$.
The induced wake structure for a supersonically (in the sense $v>u$) 
traveling color source in such a quantum liquid scenario is, quite generally, 
a conical Mach wave structure. The principal findings of our study can be 
expected to hold for any quantum liquid with a plasmon branch similiar to 
(\ref{Dispersion2}), independent of the exact form of the dielectric function.
We here assume a speed of plasmon propagation $u/c= 1/\sqrt{3}$. This differs 
from the speed of plasmon propagation in the small $k$ limit of the 
HTL approximation, $u/c=\sqrt{3/5}$ \cite{Weldon:1982aq}, and is also
different from the sound velocity in a hadronic resonance gas 
$u/c \approx \sqrt{0.2}$ \cite{Shuryak:1972zq,Venugopalan:1992hy}. 

Determining the plasmon mode via Eq.~(\ref{Dispersion2}) reveals that 
the mode is in the space-like region of the $\omega-k$ plane 
for $k>k_s=\omega_p/\sqrt{c^2-u^2}$ . Recall that this is different from the 
high-temperature plasma, where longitudinal and transverse plasma modes 
only appear in the time-like region, $|x|=|\omega/k|>1$.  In the quantum 
liquid scenario one can expect that the modes with low phase velocity 
$|x|<u/c$ suffer severe Landau damping because they accelerate the slower 
moving charges and decelerate those moving faster than the wave. 
A charge moving with a velocity lower 
than the speed of plasmon propagation can only excite those modes and 
not the modes with intermediate phase velocities $u/c<|x|<1$, which are 
undamped \cite{Ichimaru,Weldon:1982aq}. In this case, the qualitative 
properties of the color wake are analogous to those of the high temperature 
plasma: the charge carries a localized screening color cloud with it 
and Cherenkov emission and Mach cones are absent. Using (\ref{charge2}) 
and restricting the integration area to the region $k<k_c=2\omega_p$,
one can illustrate this for a colored particle traveling with $v=0.55c<u$. 
Figure \ref{figure2} shows the charge density cloud traveling with the 
colored parton.

If the colored parton travels with a velocity $v$ that is higher than the 
speed of plasmon propagation $u$, modes with an intermediate phase velocity 
$u/c<|x|<1$ can be excited. The emission of these plasma oscillations induced 
by supersonically traveling particles is analogous to Cherenkov radiation,
but different in that the density waves are longitudinal, not transverse,
excitations of the color field. Figure \ref{figure31} for a color charge
traveling with $v/c=0.99$ clearly exhibits the emergence of Mach cones in 
the induced charge density with an opening angle given by the Mach relation 
(\ref{Mach}).

\begin{figure}
\centerline{\includegraphics[width=1\linewidth]{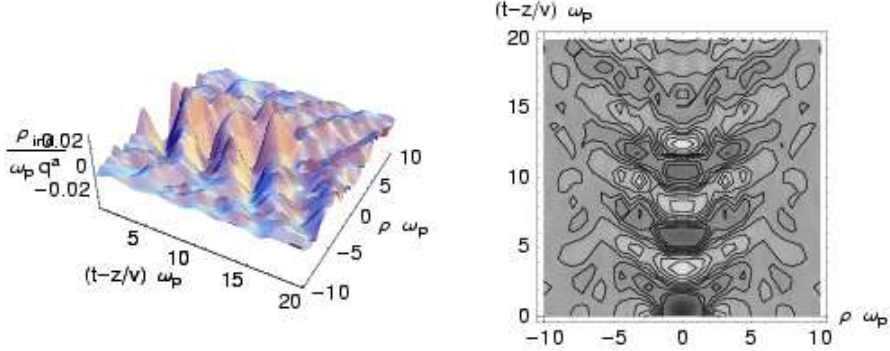}}
\caption{(a) Spatial distribution of the induced charge density from a jet 
         with high momentum and fixed color charge $q^a$ that is traveling 
         with $v=0.99c>u=\sqrt{1/3}c$. (b) Plot showing equi-charge lines 
         in the density distribution for the situation in (a).
\label{figure31}} 
\end{figure}

We emphasize that the existence of Mach cones is expected in a plasma in 
general if the particle is moving faster than the speed of sound in the 
plasma and if the dispersion relation of the collective mode extends into 
the space-like region. The wake induced by a colored jet in such a setting 
leads to regions of enhanced and depleted charge density in the wake, which 
have the shape of Mach waves trailing the projectile.

\section{Summary and Outlook}

We have calculated the properties of the color charge density wake of a hard 
parton traveling through a quark-gluon plasma in linear response theory for 
two different scenarios: a weakly coupled QGP at $T \gg T_c$ described in 
the HTL approximation, and a strongly coupled QGP with the properties of a 
quantum liquid. We found that the wake in the weakly coupled plasma always
takes the form of a screening cloud traveling with the particle, while the 
wake in the strongly coupled plasma assumes the form of a Mach cone, if the 
parton's velocity exceeds the speed of plasmon propagation and the collective 
plasma mode has a dispersion relation extending into the space-like region. 

In general, secondary particle distributions can be used to probe the 
collective excitations QCD plasma. If these take the form of Mach cones 
they should reveal themselves by the directed emission of secondary 
particles from the plasma, similar to what was observed with metallic 
targets under bombardment by swift ions. If the scenario of a strongly 
coupled QCD plasma is realized in relativistic heavy ion collisions, 
one can expect to observe these cones in the angular distribution of 
secondary particles associated with jets 
\cite{Stoecker:2004qu,Casalderrey-Solana:2004qm}. As we already mentioned
in the introduction, preliminary data from the RHIC experiments (see e.~g.\ 
Fig.~1 in \cite{Rak:2005st,Adler:2005ee}) suggest that the angular 
distribution of secondary hadrons in the direction opposite to an energetic 
hadron is peaked at an azimuthal angle $\Delta \phi \neq \pi$. This is 
different from what is observed in $p+p$ collisions, where the maximum 
associated with the away-side jet is clearly located at $\Delta \phi=\pi$. 
In contrast, two maxima are located at $\Delta \phi \approx \pi  \pm 1.1$ 
in Au + Au collisions. This phenomenon suggests the presence of a Mach 
shock front traveling with the quenched away-side jet at an angle 
$\Delta\phi = \pi \pm \varphi_M = \pi \pm {\rm arccos}(u/v)$. 

If these speculations are confirmed, it will be an interesting theoretical 
problem to determine the mechanism that excites the shock front. 
Possible candidates are the collective plasma excitations 
discussed here \cite{Ruppert:2005uz,Stoecker:2004qu}, 
the ``thunder'' effect following localized directed energy
deposition by the quenched jet \cite{Casalderrey-Solana:2004qm}, or simply
the knock-on process due to elastic collisions of plasma particles with the
away-side hard parton \cite{Lokhtin:1998ya}.
\bigskip

\section*{Acknowledgements}
This work was supported in part by U.~S.~Department of Energy under grant 
DE-FG02-05ER41367. JR thanks the Alexander von Humboldt Foundation for 
support as a Feodor Lynen Fellow. BM gratefully acknowledges partial 
support from GSI to attend this Symposium.

\end{document}